\documentclass[twocolumn,showpacs,preprintnumbers,showkeys,superscriptaddress]{revtex4}
\usepackage{graphicx}
\usepackage{dcolumn}
\usepackage{bm}

\def\eqref#1{Eq.~(\ref{eq:#1})}

\begin{document}

\title {Number of Spin $I$ States  of Identical Particles}
\author{Y. M. Zhao}     \email{ymzhao@stu.jp}
\affiliation{Department of physics,  Shanghai Jiao Tong
University, Shanghai 200030, China}
\affiliation{Cyclotron Center,  Institute of Physical Chemical Research (RIKEN), \\
Hirosawa 2-1, Wako-shi,  Saitama 351-0198,  Japan}
\affiliation{Center of Theoretical Nuclear Physics, National
Laboratory of Heavy Ion Accelerator, Lanzhou 730000, China}
\affiliation{Department of Physics, Southeast University, Nanjing
210018, China}
\author{A. Arima}
\affiliation{Science Museum, Japan Science Foundation, 2-1
Kitanomaru-koen, Chiyodaku, Tokyo 102-0091, Japan}

\date{\today}

\begin{abstract}
In this paper we study the enumeration of number (denoted as
${D_I}$) of spin $I$  states for fermions in a single-$j$ shell
and bosons with spin $l$. We show that $D_I$ can be enumerated by
the reduction   from SU$(n+1)$ to SO(3). New regularities of $D_I$
are discerned.
\end{abstract}


\maketitle

The enumeration of   number  of spin $I$ states (denoted as $D_I$)
for fermions in a single-$j$ shell or bosons with spin $l$ (We use
a convention that $j$ is a half integer and $l$ is an integer) is
a very common practice in nuclear structure theory. One usually
obtains this number by subtracting the combinatorial number  of
angular momentum projection $M=I+1$ from that with $M=I$
\cite{Lawson}. More specifically, $D_I$ equals to the
combinatorial number of $M=I$ subtracted by that of $M=I+1$, where
$M=m_1 + m_2 + \cdots + m_n$, with the requirement that $m_1 \ge
m_2 \ge \cdots \ge m_n$ for bosons and $m_1 > m_2 > \cdots > m_n$
for fermions, where $n$ is the number of particles  (This
procedure is called Process A in this paper.).  The combinatorial
numbers of different $M$'s look irregular, and such an enumeration
would be prohibitively tedious when $j$ and $l$ are very large.
The number of states of a few nucleons in a single-$j$ shell is
usually tabulated in textbooks,  for sake of convenience.

Another well-known  solution  was given by Racah \cite{Racah} in
terms of the seniority scheme, where one has to introduce (usually
by computer choice) additional quantum numbers. More than one
decade ago, a third route  was studied  by Katriel et al.
\cite{Katriel} and Sunko et al. \cite{Sunko}, who constructed
generating functions of the number of states for fermions in a
single-$j$ shell or bosons with spin $l$.

There were two efforts in constructing analytical formulas of
$D_I$. In Ref. \cite{Ginocchio}, $D_0$ for $n=4$ was obtained
analytically.  In Ref. \cite{Zhao}, $D_I$ was constructed
empirically  for   $n=3$ and 4, and some $D_I$'s for $n$=5. It is
therefore desirable to obtain a deeper insight into this difficult
problem.

Equivalent to Process A, we propose here another procedure, called
 process B  and explained  as follows. Let ${\cal P}(n, I_0)$ be the number of
partitions of $I_0 = i_1 + i_2 + \cdots i_n$, with  $0 \le i_1 \le
i_2 \le \cdots \le i_n \le 2j+1-n$ for fermions or $0 \le i_1 \le
i_2 \le \cdots \le i_n \le 2l$ for bosons. Here $I_{\rm max} = nj
-\frac{n(n-1)}{2}$ for fermions in a single-$j$
 shell, and $I_{\rm max} = nl$ for bosons with spin $l$.
One defines ${\cal P}(n, 0)= D_{I= I_{\rm max}} = 1$ for $I_0=0$.
Then one has $D_{I=I_{\rm max} - I_0} = {\cal P}(n, I_0) - {\cal
P}(n, I_0-1)$.

Now we look at $D_I$ for $\bar{n}$ ``bosons" of spin
$L=\frac{n}{2}$, with $\bar{n}=2l$ for bosons or $\bar{n}=2j+1-n$
for fermions. $I_{\rm max}$ of these $\bar{n}$ ``bosons" with spin
$L$  equals   that of $n$ bosons with spin $l$ or that of $n$
fermions in a single-$j$ shell. Furthermore,   ${\cal P}(n, I_0)$
of $I_0 = i_1 + i_2 + \cdots i_{\bar{n}}$  with the requirement $0
\le i_1 \le i_2 \le \cdots \le i_{\bar{n}} \le 2L=n$, always
equals  that  of $I_0 = i_1 + i_2 + \cdots i_n$ with the
requirement that $0 \le i_1 \le i_2 \le \cdots \le i_n \le 2j+1-n$
for $n$ fermions or  $0 \le i_1 \le i_2 \le \cdots \le i_n \le 2l$
for $n$ bosons.  This result can be explained from the fact as
follows. The ${\cal P}(n, I_0)$ of $\bar{n}$ ``bosons" with spin
$L$ corresponds to Young diagrams up to $n$ rows, and $2l$ columns
for bosons or $2j+1-n$ columns for fermions. The conjugates of
these Young diagrams are those up to $2l$ rows for bosons or
$2j+1-n$ rows for fermions, and up to $n$ columns, which
correspond to partitions in Process B for $n$ fermions in a
single-$j$ shell or bosons with spin $l$. Therefore, Process B for
$\bar{n}$ bosons with spin $L=n/2$ provides us with an alternative
to construct $D_I$ for $n$ bosons with spin $l$ or $n$ fermions in
a single-$j$ shell.

This alternative (Process B for $\bar{n}$ bosons with spin $L$)
suggests the following identity.    If $l=(2j+1-n)/2$ ($n$ is
even), i.e., $I_{\rm max}$ of bosons equals   that of fermions,
then $D_I$ for bosons equals   that of fermions. This identity can
be easily confirmed. It  means that one can obtain $D_I$ of $n$
fermions in a single-$j$ shell by using that of $n$ bosons with
spin $l=(2j+1-n)/2$, or {\it vice versa}.

Process B for $\bar{n}$ bosons with spin $L=n/2$ is also useful in
constructing formulas of $D_I$. One can see this point from the
fact that Process B involves SU($n$+1) symmetry, which is {\it
independent} of  $j$ and $l$, while  in Process A different $j$
shell for fermions and spin $l$ for bosons involve different
symmetries (SU($2j+1)$ and SU($2l+1$)~).

Below we exemplify our idea by $n=4$. The relevant symmetry for
Process B of $\bar{n}$ bosons with spin $L$ is SU(5) (i.e.,
$L=n/2=2$, $d$ bosons). $\bar{n}$ equals $2l$ and $2j-3$, for four
bosons and four fermions, respectively.

Our first result is that $D_I$ of four bosons with spin $l$ always
equals that of four fermions in a single $j$ shell when 
$l=(2j-3)/2$. Our second result is that we can derive $D_I$ of
four bosons with spin $l$ by this new method. Here one needs $D_I$
of $d$ bosons with $\bar{n}=2l$.  This problem was studied in the
interacting boson model, suggested by Arima and Iachello
\cite{Arima-Iachello} in seventies. Below we revisit the
enumeration of $D_I$  for $d$ bosons with particle number
$\bar{n}=2l$.

Let us follow the notation of Ref. \cite{Arima-Iachello} and
define $\bar{n}=2l = 2 \nu + v = 2 \nu + 3 n_{\delta} + \lambda$.
 $D_I$ of $\bar{n}$ $d$ bosons is enumerated via the procedure as
 follows.  ~ (1) ~ $v$ takes value  $2l, 2l-2, 2l-4, \cdots, 0$, which
 corresponds to $\nu = 0, 1, 2, \cdots,  n/2=l$, respectively.
  ~ (2) ~ For each value of $v$, $n_{\Delta}$ takes
 value from 0 to $\left[ \frac{v}{3} \right]$.
 ~ (3) ~ For each set of $v$ and $n_{\Delta}$,  $\lambda$
 is determined by $v-3 n_{\Delta}$.
 ~ (4) ~ For each  $\lambda$ obtained in step (3), the allowed
 spin is given by $\lambda$, $\lambda+1, \lambda+2$, $\cdots$,
 $2 \lambda-3$,  $2 \lambda-2$,  $2 \lambda$.  Note that
 there is no state with $2 \lambda-1$.
One easily sees that there is no $I=1$ states for $d$ bosons,
because $\lambda = 1$ presents $I=2$ state ($2 \lambda-1$ is
missing).

In order to obtain $D_I$, it is necessary to know the number of
$\lambda$ appearing in the above process for each $I$. Let us call
this number $f_{\lambda}$ and
  define $\bar{n}=2l=6 k + \kappa$, $\kappa = 0,   2, 4$,
and $k\ge 1$.  Below we exemplify how we obtain $f_{\lambda}$ by
the case of $\kappa=0$.  We have the following hierarchy:
\begin{eqnarray}
\lambda & f_{\lambda} & v \nonumber \\
0 & k+1 & 0,  6, 12, \cdots, 6k \nonumber \\
1 & k   & 4,  10, 16, \cdots, 6k-2 \nonumber \\
2 & k   & 2,  8, 14, \cdots,  6k-4 \nonumber \\
3 & k   & 6,  12, 18, \cdots, 6k \nonumber \\
4 & k   & 4,  10, 16, \cdots, 6k-2  \nonumber \\
5 & k-1   & 8, 14, 20, \cdots,  6k-4 \nonumber \\
6 & k   & 6,  12, 18, \cdots, 6k \nonumber \\
7 & k-1 & 10, 16, 22, \cdots, 6k-2  \nonumber \\
8 & k-1 &  8, 14, 20, \cdots,  6k-4 \nonumber \\
9 & k-1 & 12, 18, 24, \cdots, 6k  \nonumber \\
10& k-1 & 10, 16, 22, \cdots, 6k-2 \nonumber \\
11& k-2 & 14, 20, 26, \cdots,  6k-4 \nonumber \\
12& k-1 & 12, 18, 24, \cdots, 6k   \nonumber \\
13& k-2 & 16, 22, 28, \cdots, 6k-2   \nonumber \\
14& k-2 & 14, 20, 26, \cdots,  6k-4 \nonumber \\
15& k-2 & 18, 24, 30, \cdots, 6k  \nonumber \\
16& k-2 & 16, 22, 28, \cdots, 6k-2  \nonumber \\
17& k-3 & 20, 26, 32, \cdots,  6k-4  \nonumber \\
\vdots & \vdots  & ~~~~~~ ~~~~~~ \vdots
~~~~~~~~~~~~~~~~~~~~.\nonumber
\end{eqnarray}
From this tabulation we  have that $f_{\lambda} = k+ \delta_{m0} -
\delta_{m5} - \left[ \frac{\lambda}{6} \right]$, where $m$ is
equal to $\lambda$ mod 6 when $\kappa = 0$, and $\left[ ~ \right]$
means to take the largest integer not exceeding the value inside.

For the sake of simplicity  we define $I = 2 I_0$ for even values
of $I$ and $I= 2 I_0 +3$ for odd values of $I$. For $I_0 \le l$,
\begin{eqnarray}
& & D_{I=2I_0}  = \sum_{\lambda = I_0}^{2I_0} f_{\lambda}.
\end{eqnarray}

For $\kappa = 0$ and $I_0  \le l$ ($I=2I_0 \le 2l$),
\begin{eqnarray}
& & D_{I=2I_0}
 = (I_0 +1) k  \nonumber \\
&& - \left(9 K^2 - K + 3K {\cal K} +  (2 {\cal K}-5) \theta(2
{\cal K}-5)  \right) + \delta_{{\cal K}0}~, \nonumber \\
\end{eqnarray}
where $K = \left[\frac{I_0}{6} \right]$, ${\cal K} = (I_0$ mod 6).
$\theta(x) = 1$ if $x > 0$ and zero otherwise. One can repeat the
same procedure for $\kappa =   2 $ and 4. We list these results as
below:

 For $\kappa=2$ and $I_0  \le l$,
\begin{eqnarray}
&&  D_{I=2I_0}  =(I_0 +1) k  \nonumber \\
&&- \left(9 K^2 - K + 3K {\cal K}
+  (2 {\cal K}-5) \theta(2 {\cal K}-5)  \right)  \nonumber \\
& & ~~~~~ +   \left[ \frac{I_0 +3}{6} \right] +  \left[ \frac{I_0
+5}{6} \right] + \delta_{{\cal K}0}   - \delta_{{\cal K}3} ~.
\end{eqnarray}

For $\kappa=4$ and $I_0  \le  l$,
\begin{eqnarray}
&&  D_{I=2I_0}  = (I_0 +1) (k+1)  \nonumber \\
&& - \left(9 K^2 - K + 3K {\cal K}
+   (2 {\cal K}-5) \theta(2 {\cal K}-5)  \right)  \nonumber \\
&& ~~~~~ - \left[ \frac{I_0 +3}{6} \right] - \left[ \frac{I_0
+4}{6} \right]
 + \delta_{{\cal K}4}  ~.
\end{eqnarray}

For $I$ is odd and $I\le 2l$, we use a relation $D_{I=2 I_0} -
D_{I=2 I_0 +3} = \left[\frac{I_0}{2} \right] + 1$. This relation
was obtained empirically in Ref. \cite{Zhao} and can be obtained
mathematically by calculating
\begin{eqnarray}
& & D_{I=2I_0 +3} = \sum_{\lambda = I_0+3}^{2I_0+3} f_{\lambda} ~
\nonumber
\end{eqnarray}
and compare with $D_{I=2I_0}$.

For the case with $I \ge 2l$, we define $I = I_{\rm max} - 2 I_0$
for even $I$ and $I = I_{\rm max} - 2I_0 -3$ for odd $I$.
$f_{\lambda = I_0} = \left[ \frac{I_0}{6} \right] - \delta_{(I_0
{\rm ~ mod ~} 6), 0}$. We obtain that
\begin{eqnarray}
&& D_{I_{\rm max} - 2 I_0} = D_{I_{\rm max} - 2 I_0-3} = 3 \left[
\frac{I_0}{6} \right] (\left[ \frac{I_0}{6} \right]+1)
- \left[ \frac{I_0}{6} \right] \nonumber \\
& + & (\left[ \frac{I_0}{6} \right] + 1) \left( (I_0 {\rm ~ mod ~}
6) +1 \right) + \delta_{(I_0 {\rm ~ mod ~} 6), 0} -1 ~.
\end{eqnarray}

 Thus we solve the problem of enumeration of $D_I$ for four bosons
 with spin $l$ or four fermions in a single-$j$ shell by using the
  new enumeration procedure. One  may obtain $D_I$ of other $n$ ($n$ is even) cases
 by applying this method similarly, if the reduction rule of
 SU$(n+1)\rightarrow$SO(3) is available.

 A question arises when we apply this method to  odd $n$ cases, for
 which  spin $L$ of $\bar{n}$ bosons involved in Process B is not an
 integer ($L=n/2$).  These bosons are therefore not ``realistic".
 For such cases $I$ of $n$ bosons
 with spin $l$ cannot  equal  that of $n$ fermions in a
 single $j$ shell. Namely, there is no similar correspondence
 of $D_I$ between bosons and fermions when $n$ is odd  \footnote{A
correspondence of $D_I$ was noted in Sec. II of Ref. \cite{Zhao}
for large $I$ cases.}. However, $D_I$ of $\bar{n}$ fictitious
bosons with spin $n/2$  ($n$ is odd) obtained by Process A {\it
equals}  that  of $n$ bosons with spin $l$ or that of $n$ fermions
in a single-$j$ shell, where $\bar{n}=2l$ (even value) and
$2j+1-n$ (odd value) for bosons and fermions, respectively.  In
other words, $D_I$ of  $\bar{n}$ fictitious bosons with spin $n/2$
equals that of $n$ bosons with spin $l$ if $\bar{n}=2l$ or that of
$n$ fermions in a single-$j$ shell if $\bar{n}=2j+1-n$, here $n$
is odd.  Further discussion is warranted on this problem.

To summarize, We have presented in this paper an alternative to
enumerate  the number of spin $I$ states, $D_I$,  for $n$ fermions
in a single $j$ shell or $n$ bosons with spin $l$.  We proved that
$D_I$ of $n$ bosons with spin $l$ equals that of $n$ fermions in a
single-$j$ shell when $2l=2j+1-n$, where $n$ is even. We have also
exemplified the usefulness of this new method in constructing
analytical formulas of $D_I$ by $n=4$.

For odd $n$,  the procedure of our new method  involves half
integer spin $L$ for ``bosons". Further consideration of this
fictitious  situation is necessary.

We would like to thank Professors K. T. Hecht and I. Talmi  for
their reading and constructive comments of this manuscript.

\end{document}